\documentstyle[amsfonts,epsfig]{aipproc}

\begin{document}
\title{First Year Results from LOTIS}

\author{G.~G.~Williams$^1$, H.~S.~Park$^2$, E.~Ables$^2$,
D.~L.~Band$^3$, S.~D.~Barthelmy$^{4,5}$, R.~Bionta$^2$,
P.~S.~Butterworth$^4$, T.~L.~Cline$^4$, D.~H.~Ferguson$^6$,
G.~J.~Fishman$^7$, N.~Gehrels$^4$, D.~H.~Hartmann$^1$, 
K.~Hurley$^8$, C.~Kouveliotou$^7$, C.~A.~Meegan$^7$, L.~Ott$^2$,
E.~Parker$^2$, and R.~Wurtz$^2$}
\address{$^1$Dept. of Physics and Astronomy, Clemson University,
Clemson, SC 29634-1911\\
$^2$Lawrence Livermore National Laboratory, Livermore, CA 94550\\
$^3$CASS 0424, University of California, San Diego, La Jolla,
CA 92093\\
$^4$NASA/Goddard Space Flight Center, Greenbelt, MD 20771\\
$^5$Universities Space Research Association\\
$^6$Dept. of Physics, California State University at Hayward,
Hayward, CA 94542\\
$^7$NASA/Marshall Space Flight Center, Huntsville, AL 35812\\
$^8$Space Sciences Laboratory, University of California,
Berkeley, CA 94720-7450}

%\lefthead{LEFT head}
%\righthead{RIGHT head}
\maketitle

\begin{abstract}
LOTIS (Livermore Optical Transient Imaging System) is a 
gamma-ray burst optical counterpart search experiment
located near Lawrence Livermore National Laboratory in 
California.  The system is linked to the GCN (GRB Coordinates
Network) real-time coordinate distribution network and can
respond to a burst trigger in 6--15 seconds.  LOTIS has a 
total field-of-view of $17.4^{\circ} \times 17.4^{\circ}$ with
a completeness sensitivity of $m_{\rm V} \sim 11$ for a 10 second 
integration time.  Since operations began in October 1996, 
LOTIS has responded to over 30 GCN/BATSE GRB triggers.  
Seven of these triggers are considered {\em good} events 
subject to the criteria of clear weather conditions, $< 60$ s 
response time, and $> 50$\% coverage of the final BATSE 3$\sigma$ 
error circle.  We discuss results from the first year of 
LOTIS operations with an emphasis on the observations and 
analysis of GRB~971006 (BATSE trigger 6414). 
\end{abstract}

\section*{Introduction}
Our knowledge of the nature of gamma-ray bursts (GRBs)
has greatly increased as a result of recent detections 
of X-ray, optical, and radio counterparts.  X-ray 
observations of several GRBs including but not limited
to GRB~970228 and GRB~970508 by BeppoSax 
\cite{ggwilli:costa97,ggwilli:piro97} and GRB~970828 
by XTE/ASCA \cite{ggwilli:murakami97} have provided precise 
localizations which have allowed for deep optical 
follow-up searches.  These searches have resulted in the 
identification of two GRB optical counterparts, namely 
GRB~970228 \cite{ggwilli:vanparadijs97} and 
GRB~970508 \cite{ggwilli:djorgovski97} and a single radio
counterpart, GRB~970508 \cite{ggwilli:frail97}.  
Despite the wealth of information that has been obtained 
from these discoveries, the physical mechanisms which cause 
a gamma-ray burst remain a mystery.  The lack of a bright 
host galaxy for either optical counterpart has further confused 
matters.  Although the identification of the ``Hurley 100''
\cite{ggwilli:hurley97} may pin down the nature of the 
afterglows, multiwavelength observations of
GRBs {\em simultaneous} with the gamma-ray emission may
be a more direct method of probing their origin.  
If the physical processes which 
produce the prompt gamma-ray emission and the lower energy 
afterglow differ, as Katz and Piran \cite{ggwilli:katz97} 
have suggested, a broad band spectrum revealing the nature 
of the source environment can only be produced from 
simultaneous observations.   Small, wide field-of-view 
telescopes, such as LOTIS, which were originally designed 
to provide more precise burst locations by detecting the 
simultaneous optical emission may assist in producing or 
constraining this broad band spectrum. 

\section*{Observations}
LOTIS is a second generation simultaneous optical counterpart
search experiment.  The precursor experiment, called
Gamma-Ray Optical Counterpart Search Experiment (GROCSE),
found no evidence of simultaneous optical activity brighter
than m$_{\rm V}=7.5$ \cite{ggwilli:park97a}.

\begin{table}[h]
\caption{LOTIS GRB Triggers}
\label{T:ggwilli:1}
\begin{tabular}{ccddddcc}
Trig & UTC Date & Fluence/10$^{-6}$& Stat Error & Hunt-GCN Error & t$_{res}$ & Duration  \\
     &          & (erg cm$^{-2}$)    &     (deg)       &   (deg)        & (sec)     & (sec)\\
\tableline
5634 & 961017 & 0.51  & 2.9  & 2.7   & 11   & 3   \\
5719 & 961220 & 1.8   & 1.5  & 3.6   & 9    & 15  \\
6100 & 970223 & 48.0  & 0.73 & 2.0   & 11   & 30  \\
6117 & 970308 & 0.81  & 5.8  & 13.6  & 14   & 2   \\
6307 & 970714 & 1.7   & 2.8  & 7.1   & 14   & 1   \\
6388 & 970919 & 2.3   & 3.0  & 5.1   & 12   & 20  \\
6414 & 971006 & 9.3   & 0.6  & 6.8   & 17   & 150 \\
\end{tabular}
\end{table}
    
The LOTIS telescope, located 25 miles East
of Livermore, CA, consists of four individual cameras arranged
in a $2 \times 2$ array.  Each camera has a field-of-view of 
$8.8^\circ \times 8.8^\circ$ which yields a total 
field-of-view of $17.4^\circ \times 17.4^\circ$ allowing for 
a $0.2^\circ$ overlap.  A detailed description of the 
system is provided in Park {\it et al.} \cite{ggwilli:park97b} 
as well as at http://hubcap.clemson.edu/$\sim$ggwilli/LOTIS/. 
\begin{figure}[t]
\centerline{\epsfig{file=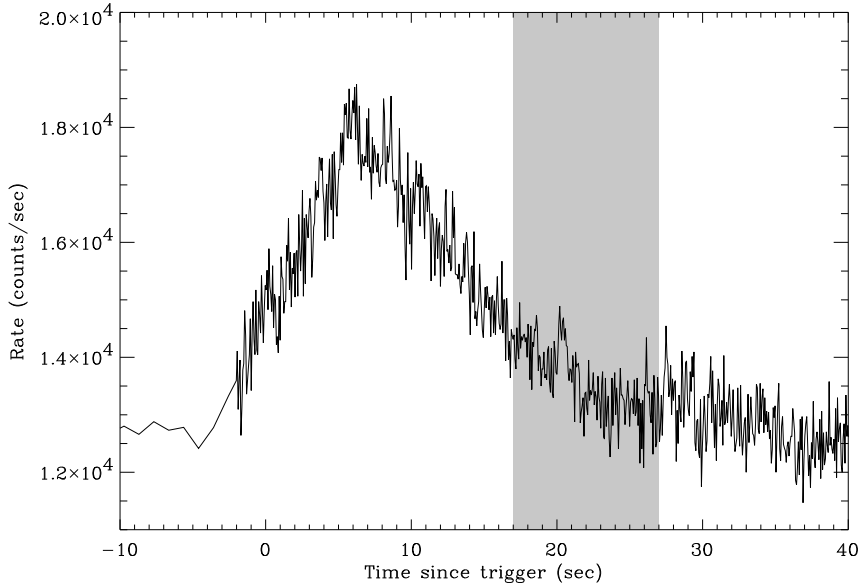,height=3.25in}}
\vspace{10pt}
\caption{GRB~971006 gamma-ray light curve.  The shaded area represents
the integration time of the first LOTIS image}
\label{F:ggwilli:1}
\end{figure}

From the start of routine operation in early October 1996 
through early October 1997 LOTIS has responded to 36 
GCN/BATSE GRB triggers \cite{ggwilli:barthelmy97}.  
Six of these triggers were a result 
of particle events occurring in the BATSE detectors.  Two 
triggers were caused by known soft gamma-ray repeaters (SGRs).  
Of the remaining 28 triggers 26 were unique GRBs while two were 
refined coordinates of previously triggered GRBs (GCN LOCBURST).
Of the 26 unique triggers 7 were considered {\em good} 
events subject to the criteria of clear weather conditions, 
$< 60$ s response time, and $> 50$\% coverage of the final 
BATSE 3$\sigma$ error circle.  By far the hardest 
criterion to meet was the coverage criterion owing to
the difficulty in determining accurate GRB locations from
the first few seconds of gamma-ray emission. Based on these
statistics LOTIS responds to $\sim 1$ good GRB event every 52 days. 

Data from the seven good events which LOTIS responded to is given in 
Table \ref{T:ggwilli:1}.  The fluence values were determined by
summing the fluence in the four energy channels given in the 
current BATSE GRB catalog with the exception of triggers 6100 and
6414 which are discussed below. The fourth column gives the 
statistical error in the final BATSE postion while the fifth 
column gives the angular difference between the initial and 
final BATSE coordinates.  The LOTIS response time and the total 
duration of the burst are given in the last two columns.  In four 
cases LOTIS began imaging while gamma-rays were still being 
emitted making the observations truly simultaneous.  
\begin{figure}[t]
\centerline{\epsfig{file=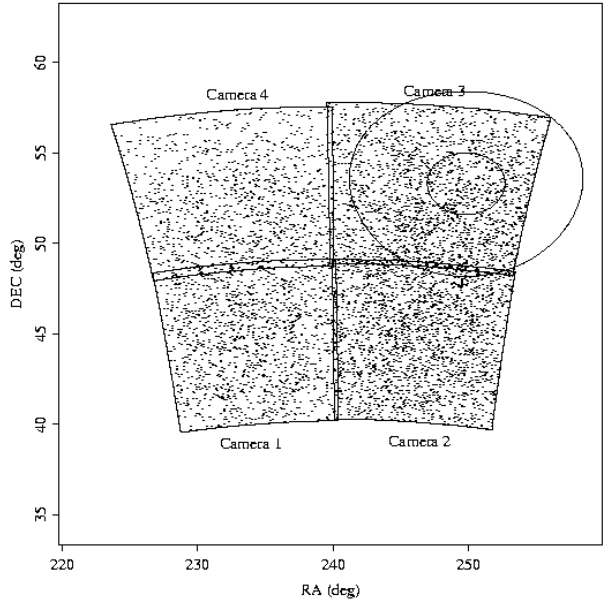,height=3.25in}}
\vspace{10pt}
\caption{LOTIS coverage of GRB~971006.  Each of the individual
points represent a stellar object above the 4$\sigma$ threshold. 
The small and large ellipses represents the BATSE 1$\sigma$ and
3$\sigma$ error circles respectively.  LOTIS obtained approximately
75\% coverage of the BATSE 3$\sigma$ error circle.}
\label{F:ggwilli:2}
\end{figure}

The event with the largest gamma-ray fluence was GRB~970223 
which was among the top 3\% of all
BATSE GRBs.  Although no optical transients were 
identified for this burst LOTIS placed an upper limit on the 
ratio of optical flux at 700 nm to gamma-ray flux at 100 keV of
$R_{F_{\rm simultaneous}}{\rm (t = 11-21 s)}=F_{\rm opt:700nm}/
F_{\gamma{\rm :100keV}} < 475$  
and on the ratio of optical to gamma-ray fluence of
$R_L=L_{\rm opt:500-850nm}/L_{\gamma{\rm :20-2000keV}} < 1.1 \times 10^{-4}$.
The full analysis of this event is given in Park {\em et al.} 
\cite{ggwilli:park97b}.

The longest burst which LOTIS responded to was GRB~971006. 
This burst had a main pulse duration of $\sim 30$ s but 
exhibited weak pre- and post-burst emission resulting in a 
total duration of $\sim 150$ s.  The light curve of 
GRB~971006 is shown in Figure~\ref{F:ggwilli:1}.
LOTIS began imaging the field centered on the initial GCN 
coordinates (RA = 241.1, Dec = 49.2 J2000) $\sim 17$ s after the 
start of the burst.  The shaded region of Figure~\ref{F:ggwilli:1}
represents the 10 s integration time of the first LOTIS image.  
The final BATSE coordinates of GRB~971006 (RA = 249.8, Dec = 53.3) 
were well within the LOTIS field-of-view.  Figure~\ref{F:ggwilli:2}
shows the LOTIS coverage for this burst.  
The small and large ellipses represent the BATSE 
1$\sigma$ and 3$\sigma$ error circles (including the 1.6$^\circ$ 
systematic error) respectively.  There was no  
Interplanetary Network (IPN) \cite{ggwilli:hurley94} localization available 
for this burst and therefore it was necessary to search the entire 
3$\sigma$ error circle for transient objects.  This search 
found no transients with a point spread function (psf) consistent 
with the stellar psf.  

From a histogram plot of stellar 
magnitudes in camera 3 we determined the completeness magnitude 
(the faintest magnitude for which 100\% of the stars were detected)
for this event to be m$_{\rm V} \sim 11.0$.  
Following the analysis in Park {\em et al.} \cite{ggwilli:park97b}
this yields an upper limit to 
the flux density at 700 nm of 
$F_{\nu}{\rm (700 nm)}<2.7 \times 10^{-24}$ erg~cm$^{-2}$~s$^{-1}$~Hz$^{-1}$.
The BATSE flux density at 100 keV was found by fitting the 
spectrum from LAD3 during the integration time of the first 
LOTIS image to the Band GRB functional form \cite{ggwilli:band93} 
which yielded a value of
$F_{\nu}{\rm (100 keV)}=1.7 \times 10^{-27}$ erg~cm$^{-2}$~s$^{-1}$~Hz$^{-1}$.
The resulting upper limit of the optical to gamma-ray flux for this
event is 
$R_{F_{\rm simultaneous}}{\rm (t = 17-27 s)}=F_{\rm opt:700nm}/
F_{\gamma{\rm :100keV}} < 1600$. 

The total gamma-ray fluence was determined by integrating the 
Band GRB functional form for the entire burst from 20 keV 
to 2000 keV.  The total gamma-ray fluence was
$L_{\gamma :20-2000{\rm keV}}=9.3 \times 10^{-6}$ erg~cm$^{-2}$ 
while the upper limit to the GRB's optical fluence, 
again following Park {\em et al.} \cite{ggwilli:park97b}, is
$L_{\rm opt:500-850nm}<5.4 \times 10^{-9}$ erg~cm$^{-2}$. 
The resulting upper limit for the optical to gamma-ray fluence 
ratio is 
$R_L=L_{\rm opt:500-850nm}/L_{\gamma{\rm :20-200keV}} < 5.8 \times 10^{-4}$.

\section*{Discussion}
Although LOTIS has already placed upper limits on the simultaneous 
optical to gamma-ray flux for specific events we hope 
to further constrain the ratio with an upgrade to 
thermo-electric cooled CCDs in January 1998.  In the future 
we plan to investigate GRB spectral evolution focusing 
on how the low energy power law index, $\alpha$, of the Band GRB 
functional form \cite{ggwilli:band93} effects optical constraints. 
We also plan to implement 
Super-LOTIS \cite{ggwilli:park97c}, a dedicated 0.6 m reflector 
with a design sensitivity of m$_{\rm V} \sim 19$ 
(10 s integration time) early next year.

\end{document}